\documentclass[aps,pre,preprint,superscriptaddress]{revtex4}
\usepackage{graphicx}
\usepackage{color}
\usepackage{bm}
\usepackage{epstopdf}
\usepackage{mathtools}
\begin{document}



\title{Planar Rotational Equilibria of Two Non-identical Microswimmers }

\author{Prajitha Mottammal}

\address{Department of Chemical Engineering, Indian Institute of Technology Madras, Chennai, India}

\author{Sumesh P. Thampi}
\address{Department of Chemical Engineering, Indian Institute of Technology Madras, Chennai, India}

\author{Andrey Pototsky}
\address{Department of Mathematics, Swinburne University of Technology, Hawthorn, Victoria, 3122, Australia}


\begin{abstract}
 We study a planar motion of two hydrodynamically coupled non-identical micro-swimmers, each modelled as a force dipole with intrinsic self-propulsion. Using the method of images, we demonstrate that our results remain equally applicable at a stress-free liquid-gas interface as in the bulk of a fluid.
Closed analytical form of circular periodic orbits for a pair of two pullers and a pair of two pushers is presented and their linear stability is determined with respect to two- and three-dimensional perturbations. A universal stability diagram of the orbits with respect to two-dimensional perturbations is constructed and it is shown that two non-identical pushers or two non-identical pullers moving at a stress-free interface may form a stable rotational equilibrium.
 For two non-identical pullers we find stable quasi-periodic localized states, associated with the motion on a two-dimensional torus in the phase space. Stable tori are born from the circular periodic orbits as the result of a torus bifurcation.
All stable equilibria in two-dimensions are shown to be monotonically unstable with respect to three-dimensional perturbations. 
\end{abstract}

\keywords{low-Reynolds-number flows, self-propulsion, rotational equilibrium}
\maketitle
\section{Introduction}
\label{intro}
Hydrodynamic interactions are known to play a major role in the locomotion of microswimmers such as motile bacteria, spermatozoa, or artificial self-propelled particles \cite{GUELL88,lauga2009hydrodynamics,Dunkel13,Ariel15,Sokolov2007}. In viscous fluids, the motion on the micro-scale $L\approx 1-10 \mu$m typically occurs with velocities $V$ of up to ten micrometers per second, which translates to a small Reynolds number ${\rm Re}=\rho VL/\eta \ll 1$, where $\rho$ is the density and $\eta$ is the dynamic viscosity of the incompressible fluid. Therefore, in this regime, the inertial effects are negligible and the flow perturbed by swimmer's moving body propagates instantly to all parts of the fluid removing any time lag between action and hydrodynamically mediated re-action. In addition to being instantaneous, the strength of the induced flow field decays slowly as a power law function of the distance from a swimmer, giving rise to long-range interactions \cite{Batchelor1970}.

 In the simplest case of two force-free swimmers far away from any interfaces or boundaries, the effect of hydrodynamic interactions strongly depends on the ratio of swimmer's size to the separation distance $r$, as well as on their propulsion mechanism.   Thus, two model squirmers in the near-field regime, i.e. when the separation distance  is comparable with their size, turn their bodies into each other leading to effective attraction. As the separation distance decreases, the orientations of swimmers rapidly change in such a way that the swimmers turn  away from each other and eventually separate \cite{Ishikawa2006}. Experiments conducted with two {\it Paramecia caudatum} cells on a collision course revealed that the cells avoid each other due to hydrodynamically induced re-orientation of their bodies \cite{ishikawa2006interaction}. 

In the far-field, when the distance $r$ between swimmers is much larger than their size, the leading order flow generated by a force-free swimmer is that of a force dipole which decays as $\sim 1/r^2$ \cite{Batchelor1970}. Far-field dynamics of two identical force dipoles with an intrinsic rotation of their orientations was studied in \citet{michelin2010long}. Two types of unstable rotational equilibria and one type of non-rotational equilibrium were found in the regime of fast intrinsic rotation and slow coupled dynamics of two swimmers. Due to the unstable nature of the equilibria, only two types of the long time behaviour were shown to be possible: monotonic attraction, terminated by collision, or monotonic repulsion and separation.   

In artificial swimmers whose body consists of three or more linked spheres, the swimming stroke can be designed in such a way that the force dipole contribution to the flow is absent and the leading order flow decays as $\sim 1/r^3$ \cite{pooley2007hydrodynamic,Farzin2012}. Collective motion of two such swimmers strongly depends on the initial position and the relative phase of the stroke mechanism. In the long time limit the swimmers either attract, or repel each other monotonically - similar to two force dipoles \cite{michelin2010long}, or remain oscillatory. For specific initial conditions two swimmers swimming in the same direction also exhibit a sustained oscillatory motion about a common center with a slow drift, which points towards the existence of a rotational equilibrium. Numerical simulations revealed that this oscillatory state is long-lasting, however a detailed stability analysis is still missing. For the majority of initial conditions and relative phase of the stroke mechanism, two linked-sphere model swimmers hydrodynamically scatter in such a way that their relative angle is preserved \cite{Alexander2008}.

Coupled dynamics of more complicated model-microswimmers has also been studied in the literature, for example for two hydrodynamically coupled identical Quadroars, i.e. swimmers, whose bodies consist of four linked rotary discs \cite{Mirzakhanloo2018}. Depending on their initial positions and orientations, the long-time dynamics of two Quadroars is remarkably reach, featuring non-orbiting monotonically converging or diverging paths, a pursuit-evasion type of dynamics, and capture into bound quasi-periodic orbits. Numerical simulations revealed that dynamical equilibria are stable to small perturbations.

Recently we have studied the co-planar dynamics of two identical force dipoles in the bulk of a fluid and at a fluid-gas interface in the presence of self-induced Marangoni flow \cite{MTP21}. When moving in the bulk of a fluid, the trajectories of the swimmers are two-dimensional and remain co-planar at all times, if their initial orientations and the relative position vector are co-planar.  We have derived a closed analytical form of two types of rotational equilibrium states, both characterised by a circular motion of swimmers with constant relative orientation. The two types of equilibria coexist and can be reached by an appropriate choice of the initial conditions. Linear stability analysis revealed that both types are orbitally unstable with either one real unstable Floquet multiplier or two complex-conjugate unstable multipliers. As a consequence, the swimmers escape from the equilibria either monotonically or oscillatory. We have shown that the equilibria can be stabilized by self-induced Marangoni flow, when both swimmers are chemically active and move at a planar fluid-gas interface.

Here we consider the planar motion of two non-identical swimmers with self- propulsion velocities $v_i$, modeled as force dipoles with different dipole strength $p_i$, $(i=1,2)$. Using the method of images, one can readily show that a planar motion of two force dipoles moving in the bulk of unbounded fluid is identical to the motion at a stress-free interface, provided that the orientation vectors of the swimmers are parallel to the plane of motion. In this paper we show that similar types of rotational equilibria, as found in \cite{MTP21}, also exist for a pair of two non-identical pushers or pullers. These equilibria are associated with periodic orbits of the reduced dynamical system. We use numerical continuation method \cite{AUTO} to study the orbital stability of the equilibria with respect to two- and three-dimensional perturbations. Using the ratio of the dipole strengths $p_2/p_1$ and the ratio of self-propulsion speeds $v_2/v_1$, a universal stability diagram of the periodic orbits is constructed with respect to two-dimensional perturbations. Thus, two non-identical pushers or pullers bound to move along a stress-free interface may form a stable rotational equilibrium, if their orientation vectors remain parallel to the interface at all times. In addition, for two non-identical pullers we find a quasi-periodic stable equilibrium, associated with the motion on a two-dimensional torus in the phase space. Invariant torus is born as a result of a super-critical torus bifurcation from the periodic orbit.  All orbits that are stable with respect to two-dimensional perturbations are shown to be monotonically unstable with respect to three dimensional perturbations, as characterised by a single real unstable Floquet multiplier.

\section{Planar motion of two force-free hydrodynamically coupled swimmers}
\label{orbits}
Consider two non-identical force-free swimmers moving in the bulk of a fluid with dynamic viscosity $\eta$. Each swimmer is modeled as a force dipole with the dipole strength $p_i$, orientation ${\bm e}_i$ and an intrinsic self-propulsion velocity $v_i$, $(i=1,2)$.  The position vector of the $i$-th swimmer is ${\bm r}_i$.
In the limit of vanishing inertia, the instantaneous velocity $d{\bm r}_i/dt$ of the $i$-th swimmer is a superposition of the self-propulsion velocity $v_i{\bm e}_i$ and the the flow induced by the other swimmer at the current position of swimmer $i$
\begin{eqnarray}
\label{eq1}
\frac{d{\bm r}_i}{dt}&=&v_i{\bm e}_i+\frac{p_k}{8\pi\eta}\left(3\left(\frac{{\bm e}_k\cdot {\bm r}_{ik}}{|{\bm r}_{ik}|}\right)^2-1\right)\frac{{\bm r}_{ik}}{|{\bm r}_{ik}|^3},
\end{eqnarray}
where ${\bm r}_{ik}={\bm r}_i-{\bm r}_k$ is the relative position vector.
The curl of the flow field generated by swimmer $i$ at the position of swimmer $k$ is \cite{lauga2009hydrodynamics}
\begin{eqnarray} 
\label{eq2}
{\bm \Omega}_{ik}=\frac{3p_i}{4\pi\eta}\frac{({\bm e}_i\cdot {\bm r}_{ki})({\bm e}_i\times  {\bm r}_{ki})}{|{\bm r}_{ki}|^5}.
\end{eqnarray}
Using the well-known Fax{\'e}n's law, the angular velocity of the $i$-th swimmer is given by the half of the curl of the velocity field generated by the other swimmer at the position of the $i$-th swimmer \cite{Faxen} 
\begin{eqnarray} 
\label{eq3}
\frac{d{\bm e}_i}{dt}&=&\frac{1}{2}{\bm \Omega}_{ki}\times {\bm e}_i.
\end{eqnarray}
Eqs\,(\ref{eq1},\ref{eq2},\ref{eq3}) describe the translational and rotational motion of a pair of force dipoles with intrinsic propulsion.
From Eqs\,(\ref{eq1},\ref{eq2},\ref{eq3}) it is also clear that if initially the vectors ${\bm e}_1$, ${\bm e}_2$ and ${\bm r}_{12}$ are co-planar, then the motion of the swimmers remains planar at all times.

 Eqs\,(\ref{eq1},\ref{eq2},\ref{eq3}) are equally applicable to describe the motion of two force dipoles at a planar stress-free liquid-gas interface, in the case if the orientation vectors ${\bm e}_i$ remain parallel to the interface at all times. Indeed, the flow field ${\bm v}$ 
 induced by a force dipole with the orientation vector ${\bm e}$ parallel to the interface automatically fulfills the stress-free boundary condition $\partial_z {\bm v}=0$. A more systematic way of describing the motion of swimmers close to interfaces is the method of images. Thus, consider a force dipole with strength $p$ and orientation ${\bm e}$ parallel to a stress-free interface $z=0$. If the force dipole is located at $z=-h$, then 
using the method of images \cite{lauga2009hydrodynamics} an identical force dipole must be introduced at $z=h$ to fulfill the stress-free boundary conditions at $z=0$. The motion at the interface is then recovered in the limit $h\rightarrow 0$, when the original force dipole and its image coincide.

In what follows we choose the plane $z=0$ in Cartesian coordinates $(x,y,z)$ as the plane of motion. The time-dependent position vectors of the swimmers are then given by ${\bm r}_i(t)=(x_i(t),y_i(t),0)$, with $i=1,2$ and the unit orientation vectors of the swimmers can be represented in the form ${\bm e}_i=(\cos(\phi_i),\sin(\phi_i),0)$.

Introducing the distance between swimmers $\rho=|{\bm r}_{12}|$ and the angles $\theta_{ki}$ according to  $\cos(\theta_{ki})=\rho^{-1}{\bm e}_k \cdot {\bm r}_{ik}$, we obtain from Eqs.\,(\ref{eq2},\ref{eq3})
\begin{eqnarray} 
\label{eq4}
\frac{d\phi_i}{dt}=\frac{3p_k}{16\pi \eta}\frac{\sin(2\theta_{ki})}{\rho^3}.
\end{eqnarray}

\begin{figure}[ht]
\includegraphics[width=0.9\hsize]{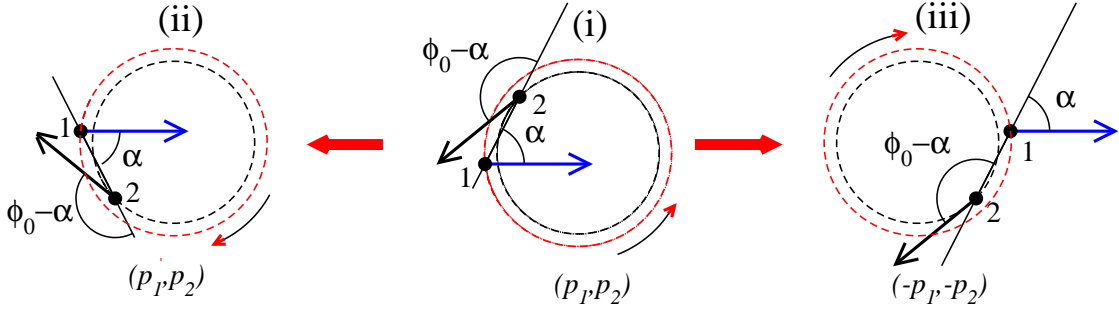} 
\caption{\label{F1}  Two types of symmetry of the rotational equilibria. If an orbit $(\rho,\omega,\alpha,\phi_0)$ exists for some combination of $(p_1,p_2)$, then $(\rho,-\omega,-\alpha,-\phi_0)$ is also an orbit for the same pair of swimmers $(p_1,p_2)$: (i)$\rightarrow$(ii) and  $(\rho,-\omega,\alpha-\pi,\phi_0)$ is also an orbit for the swimmers with opposite polarity $(-p_1,-p_2)$: (i)$\rightarrow$(iii).} 
\end{figure}
%

\subsection{Reduced dynamical system and periodic solutions}
\label{reduced}
Eqs.\,(\ref{eq1},\ref{eq4}) are invariant with respect to simultaneous translations of both swimmers, which allows us to reduce their coupled dynamics to a four-dimensional phase space, constructed using the variables $(\Delta x=x_1-x_2,\Delta y=y_1-y_2,\phi_1,\phi_2)$. 
The resulting reduced system can be written in the form
\begin{eqnarray}
\label{eq6}
\frac{d(\Delta x)}{dt}&=&v_1\cos(\phi_1)-v_2\cos(\phi_2)+\frac{E\Delta x}{8\pi\eta\rho^3}\nonumber\\
\frac{d(\Delta y)}{dt}&=&v_1\sin(\phi_1)-v_2\sin(\phi_2)+\frac{E\Delta y}{8\pi\eta\rho^3}\nonumber\\
\frac{d\phi_1}{dt}&=&\frac{3p_2}{8\pi \eta\rho^3}\cos(\theta_{21})\sin(\theta_{21}),\nonumber\\
\frac{d\phi_2}{dt}&=&\frac{3p_1}{8\pi \eta\rho^3}\cos(\theta_{12})\sin(\theta_{12}),
\end{eqnarray}
with $E=3p_2\cos^2(\theta_{21}) +3p_1\cos^2(\theta_{12})-p_1-p_2$, $\cos{\theta_{12}}=-\rho^{-1}(\Delta x \cos(\phi_1)+\Delta y\sin(\phi_1))$, $\cos{\theta_{21}}=\rho^{-1}(\Delta x \cos(\phi_2)+\Delta y\sin(\phi_2))$, $\sin(\theta_{12})=-\rho^{-1}(\Delta y \cos(\phi_1)-\Delta x\sin(\phi_1))$ and $\sin(\theta_{21})=\rho^{-1}(\Delta y \cos(\phi_2)-\Delta x\sin(\phi_2))$.

Any rotational equilibrium corresponds to periodic solutions of Eqs.\,(\ref{eq6}), modulo a phase gain of $2\pi$ for the orientation angles $\phi_1$ and $\phi_2$. Similar to \cite{MTP21}, we look for a simple harmonic rotation along a circular path with radius $\rho$ and a constant frequency $\omega$
\begin{eqnarray}
\Delta x&=& -\rho\cos(\alpha+\omega t),\nonumber\\
\Delta y&=& -\rho\sin(\alpha+\omega t),\nonumber\\
\phi_1&=&\omega t,\nonumber\\
\phi_2&=&\omega t+\phi_0,
\label{eq7}
\end{eqnarray}
where $\phi_0=\phi_1-\phi_2$ represents the relative orientation angle and $\alpha$ is the angle formed by the orientation vector of the first swimmer ${\bm e}_1=(\cos(\phi_1),\sin(\phi_1))$ and the distance between swimmers $(\Delta x,\Delta y)$.  From Eqs.\,(\ref{eq7}) we obtain $\cos{\theta_{12}}=\cos(\alpha)$, $\cos{\theta_{21}}=-\cos(\alpha-\phi_0)$,  $\sin{\theta_{12}}=\sin(\alpha)$, $\sin{\theta_{21}}=-\sin(\alpha-\phi_0)$. 

Substitution of Eqs.\,(\ref{eq7}) into Eqs.\,(\ref{eq6}) yields the solvability conditions
\begin{eqnarray}
\rho\omega&=&v_1\sin(\alpha)-v_2\sin(\alpha-\phi_0),\nonumber\\
\frac{E}{8\pi\eta\rho^2}&=&v_1\cos(\alpha)-v_2\cos(\alpha-\phi_0),\nonumber\\
\omega&=&\frac{3p_1}{16\pi \eta\rho^3}\sin(2\alpha),\nonumber\\
\omega&=&\frac{3p_2}{16\pi \eta\rho^3}\sin(2(\alpha-\phi_0)).
\label{eq7a}
\end{eqnarray}

If, for a certain combination of parameters $p_i$ and $v_i$ there exist such $\alpha$, $\phi_0$, $\omega$ and $\rho$ so that  the conditions Eqs.\,(\ref{eq7a}) are satisfied, then the rotational equilibrium, described by Eqs.\,(\ref{eq7}) exists and the coordinates of the swimmers $x_i$ and $y_i$ in the laboratory frame can be obtained by integrating Eqs.\,(\ref{eq1})
\begin{eqnarray}
x_1&=&x_0+\frac{v_1\sin(\phi_1)}{\omega}+\frac{p_2[3\cos^2(\alpha-\phi_0)-1]}{8\pi \eta\omega \rho^3}\Delta y,\nonumber\\
y_1&=&y_0-\frac{v_1\cos(\phi_1)}{\omega}-\frac{p_2[3\cos^2(\alpha-\phi_0)-1]}{8\pi \eta\omega \rho^3}\Delta x,\nonumber\\
x_2&=&x_0+\frac{v_2\sin(\phi_2)}{\omega}-\frac{p_1[3\cos^2(\alpha)-1]}{8\pi \eta\omega \rho^3}\Delta y,\nonumber\\
y_2&=&y_0-\frac{v_2\cos(\phi_2)}{\omega}+\frac{p_1[3\cos^2(\alpha)-1]}{8\pi \eta\omega \rho^3}\Delta x,
\label{eq8}
\end{eqnarray}
where $(x_0,y_0)$ is an arbitrary position of the centre of circular paths.

From Eqs.(\ref{eq8}) it is easy to obtain the radii $R_1$ and $R_2$ of the circular paths
\begin{eqnarray}
R_1^2&=&\frac{v_1^2}{\omega^{2}}+\frac{p_2^2(3\cos^2(\alpha-\phi_0)-1)^2}{(8\pi \eta \omega \rho^2)^2}-\frac{2p_2(3\cos^2(\alpha-\phi_0)-1)\cos(\alpha)}{8\pi \eta(\omega\rho)^2},\nonumber\\
R_2^2&=&\frac{v_2^2}{\omega^{2}}+\frac{p_1^2(3\cos^2(\alpha)-1)^2}{(8\pi \eta \omega\rho^2)^2}+\frac{2p_1(3\cos^2(\alpha)-1)\cos(\alpha-\phi_0)}{8\pi \eta(\omega\rho)^2}.
\label{eq8a}
\end{eqnarray}
%
%

\subsection{Pusher-pusher and puller-puller periodic orbits}
\label{sym}
Solvability conditions Eqs.\,(\ref{eq7a}) have three different types of symmetry. The first type is associated with the existence of identical orbits with either clockwise or anti-clockwise rotation.   Thus, for any $v_i$ and $p_i$, if $(\rho,\omega,\alpha,\phi_0)$ is a solution of Eqs.\,(\ref{eq7a}), then $(\rho,-\omega,-\alpha,-\phi_0)$ is also a solution for the same set of $v_i$ and $p_i$. The second type of symmetry is associated with replacing each pusher (puller) with the puller (pusher) of identical dipole strength. Indeed, if $(\rho,\omega,\alpha,\phi_0)$ is a solution of Eqs.\,(\ref{eq7a}) for a particular combination of $(p_1,p_2)$, then $(\rho,-\omega,\alpha-\pi,\phi_0)$ is also a solution for a pair of swimmers with reversed polarity, i.e. $(-p_1,-p_2)$, rotating in the opposite direction. These two types of symmetry are schematically shown in Fig.\,\ref{F1}: if orbit (i) exists for $(p_1,p_2)$, then orbit (ii) also exists for $(p_1,p_2)$ and orbit (iii) exists for $(-p_1,-p_2)$.

The third type of symmetry of periodic orbits is associated with an arbitrary choice of numbering of swimmers. Indeed, it is clear that if an orbit exists for a certain combination of $(p_1,p_2)$ and $(v_1,v_2)$, then the same orbit also exists when the swimmers are interchanged, i.e. if $1\rightleftharpoons 2 $.   From  Fig.\ref{F1}, we see that when the swimmers are interchanged (renumbered), the angles $\alpha$ and $\phi_0-\alpha$ are replaced with $\pi-(\phi_0-\alpha)$ and $\pi-\alpha$, respectively. Simultaneously the dipole strengths and the self-propulsion velocities are interchanged $p_1\rightleftharpoons p_2$ and  $v_1 \rightleftharpoons v_2$. Under the above transformations, the distance between swimmers $\rho$ and rotational frequency $\omega$ remain unchanged.

Eqs.\,(\ref{eq6}) can be non-dimensionalized by scaling the coordinates and distances with $\lambda=\sqrt{p_1/(8\pi\eta v_1)}$ and time with $\tau=\lambda/v_1$. This reduces the number of independent dimensionless parameters to two, namely, the ratio of the propulsion velocities $v=v_2/v_1$ and the ratio of the dipole strengths $p=p_2/p_1$. 
The dimensionless form of Eqs.\,(\ref{eq6}-\ref{eq8a}) is obtained by replacing $v_1\rightarrow 1$, $v_2\rightarrow v$, $p_1/(8\pi\eta)\rightarrow 1$ and $p_2/(8\pi\eta)\rightarrow p$.

\begin{figure}[ht]
\includegraphics[width=0.9\hsize]{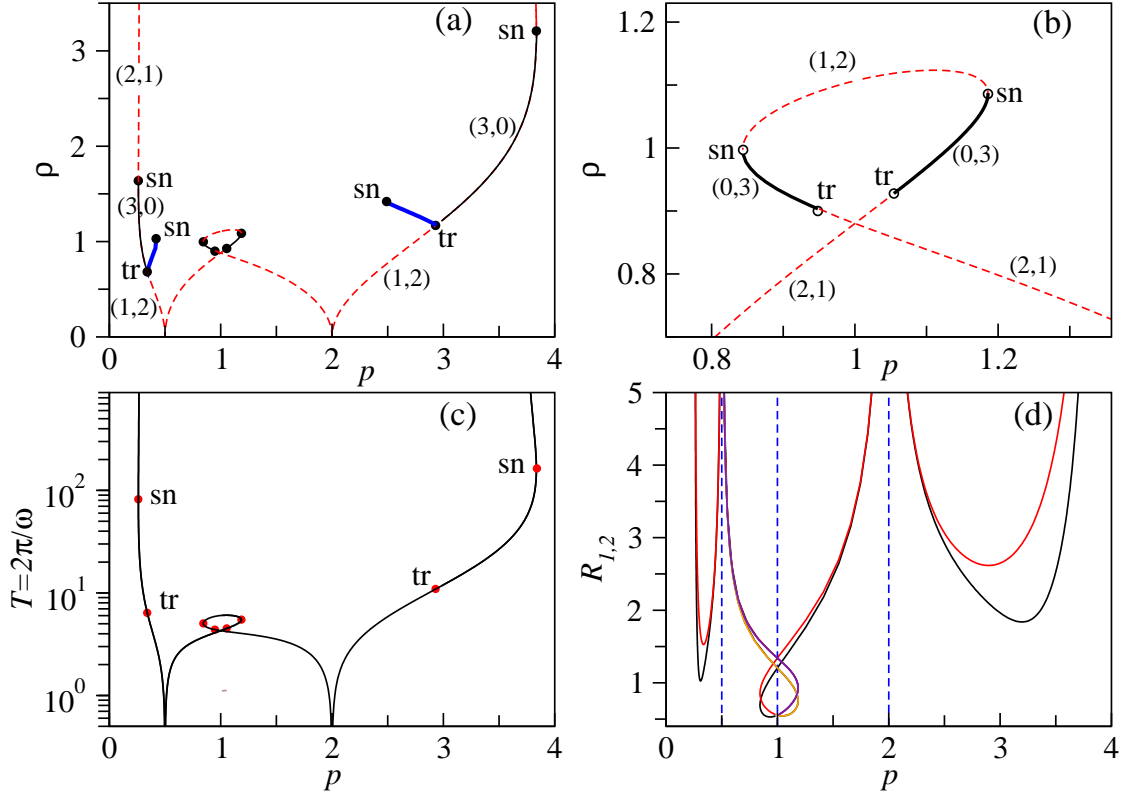} 
\caption{\label{F2} Bifurcation diagram of Eqs.\,(\ref{eq6}) for a pair of two pushers ($p_i>0$) or a pair of two pullers ($p_i<0$) with $v=1$ and $p=p_1/p_2$ as a bifurcation parameter. (a,b) Distance between swimmers $\rho$. Thin solid (dashed) lines represent orbitally stable (unstable) periodic solutions Eqs.\,(\ref{eq7}). Thick solid lines correspond to stable quasi-periodic solutions, associated with the motion on a torus (Fig.\ref{F3}). The first (second) number in the brackets near each curve shows the number of unstable Floquet multipliers of periodic solutions for a pair of two pushers (pullers).  Labels {\tt sn} and {\tt tr} correspond to the saddle-node and the torus bifurcation points, respectively. (b) Zoomed plot of the central loop in (a). Orbital period $T=2\pi/\omega$ (c) and radii $R_i$ of the orbits in the laboratory frame (d). } 
\end{figure}

Earlier, we have shown that for a pair of identical pushers, i.e. $p_1=p_2>0$, the system Eqs.\,(\ref{eq7a}) admits two solutions that can be found analytically \cite{MTP21}.  For an arbitrary combination of parameters $p=p_2/p_1$ and $v=v_2/v_1$, transcendental equations Eqs.\,(\ref{eq7a}) can only be solved numerically. Here instead of solving Eqs.\,(\ref{eq7a}), we use numerical continuation method \cite{AUTO} to directly follow the branch of periodic solutions of Eqs.\,(\ref{eq6}) modulo a phase gain of $2\pi$. The analytically known solution at $p=1$ and $v=1$, reported in \cite{MTP21}, is continued using $p$ as the principal continuation parameter. The orbital stability of the equilibria is determined automatically by numerically linearizing Eqs.\,(\ref{eq6}) about the periodic orbit and then computing the Floquet multipliers of the linearized Poincare map \cite{AUTO} . For periodic solution, one of the multipliers is always real and equal to one. The remaining (possibly complex) multipliers determine the stability of the equilibria.

First, at fixed $v=1$ we vary $p$ and find three disconnected branches of periodic solutions of Eqs.\,(\ref{eq6}), as shown in Fig.\,\ref{F2}. The distance between the swimmers $\rho$ is plotted {\it vs} $p$ in Fig.\,\ref{F2}(a,b). The central branch has a loop in the middle and terminates at points $p=0.5$ and $p=2$. At these points, the radii $R_i$ of the swimmer's trajectories in the laboratory frame, given by Eqs.\,(\ref{eq8a}), and the angular frequency $\omega$ diverge, i.e. $R_i\rightarrow \infty$ and $\omega\rightarrow \infty$, while the period of revolution $T=2\pi/\omega$ tends to zero, as shown in Fig.\,\ref{F2}(c,d).

According to the symmetry (i)$\rightarrow$(iii) in Fig.\,\ref{F1}, a pair of two pushers and a pair of two pullers with identical dipole strength ratio $p$ have identical swimmer-to-swimmer distance $\rho$ and orbital frequency $\omega$.  However, the orbital stability for a pair of pushers differs from that for a pair of pullers, because the linearized Eqs.\,(\ref{eq6}) does not have the symmetry of Eqs.\,(\ref{eq7a}).
The first (second) number in the bracket near each part of the branch in Fig.\,\ref{F1}(a,b) shows the number of the unstable Floquet multipliers for pushers (pullers). If at least one of the numbers in the bracket is zero, the corresponding periodic solution is orbitally stable. At any of the saddle-node bifurcation points (sn) one real multiplier is equal to one. Similarly, at the torus bifurcation point (tr) a pair of complex multipliers cross a unit circle.

We find stable orbits for a pair of two non-identical pushers on the central solution branch $(0.5<p<2)$ and stable orbits for a pair of two non-identical pullers on the left and right branches $(p<0.5)$ and $(p>2)$, as shown in Fig.\ref{F1}(a,b). Each orbit is a circular limit cycle in the four-dimensional phase space of Eqs.\,(\ref{eq6}). 

Due to the symmetry associated with an arbitrary numbering of swimmers $1\rightleftharpoons 2 $, the bifurcation diagram in Fig.\,(\ref{F2})(a,b) is invariant under the simultaneous transformations $p\rightarrow 1/p$ followed by the re-scaling of the dimensionless distance between swimmers $\rho$ according to $\rho\rightarrow \rho \sqrt{p}$. This implies that when the self-propulsion velocities of the swimmers are identical, i.e. $v_1=v_2$,  the solution branches that exist for $p<1$ are equivalent to those found for $p>1$.  
However, for consistency of presentation, we select to show both parts of the bifurcation diagram in Fig.\,(\ref{F2}), taking into account that in what follows (Fig.\,(\ref{F5})(a)) we also discuss the fully asymmetric case, when $v_1\not=v_2$ and $p_1\not=p_2$.

\begin{figure}[ht]
\includegraphics[width=0.9\hsize]{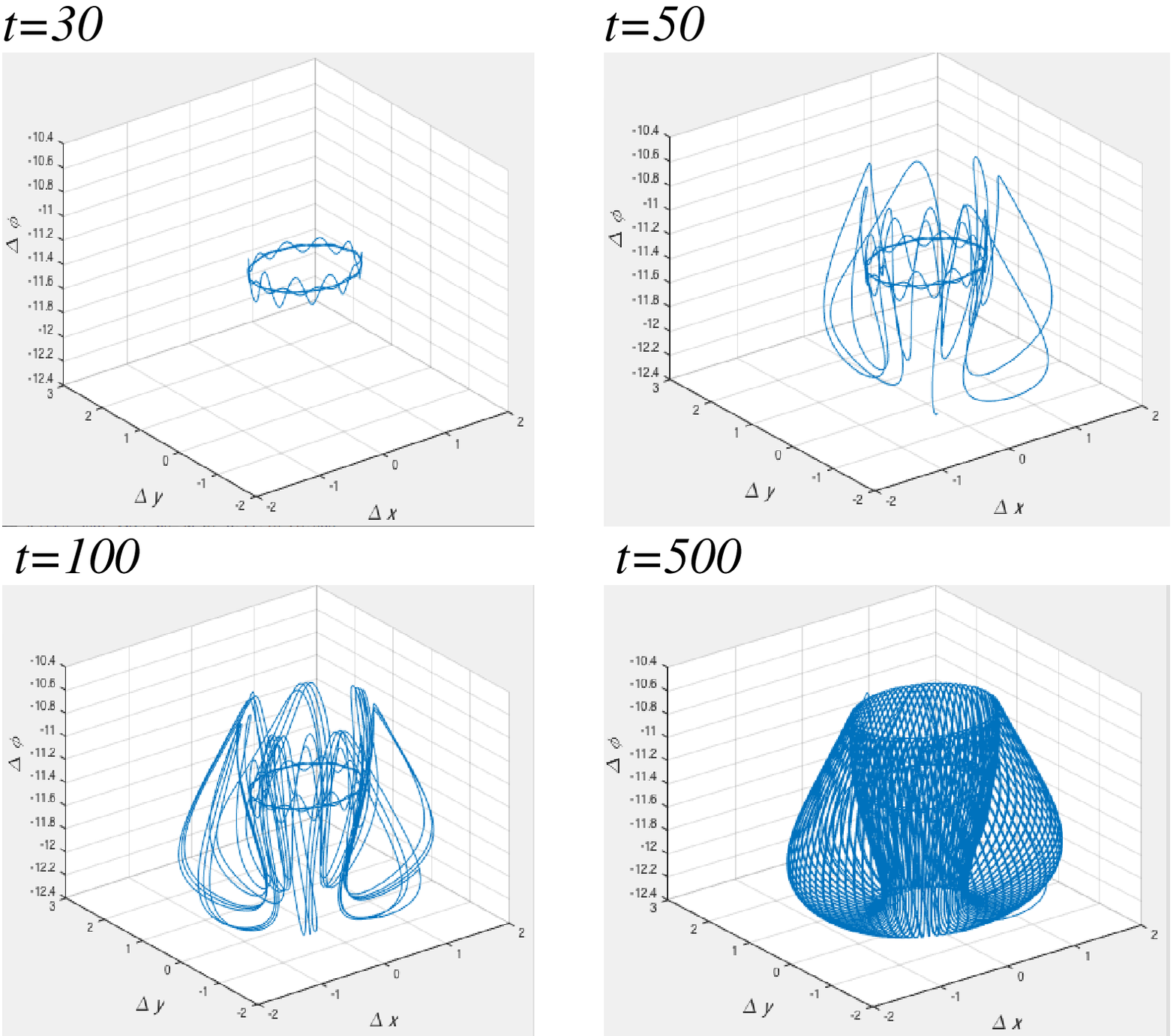} 
\caption{\label{F3} Formation of a stable two-dimensional torus for a pair of two pullers with $p=2.5$ and $v=1$. Initial conditions are chosen close to the unstable limit cycle. } 
\end{figure}
\subsection{Quasi-periodic orbits}
\label{quasi}
For parameters on the central solution branch $(0.5<p<2)$, the motion around any unstable orbit is either unbounded, leading to an infinite separation between swimmers ($\rho\rightarrow \infty$), or terminal, associated with a finite-time collision  ($\rho\rightarrow 0$). 
However, near the torus bifurcation points on the left $p<0.5$ and on the right $p>2$ branches we find stable bounded states for a pair of two non-identical pullers, which can be associated with the motion on a two-dimensional torus.

 The formation of the two-dimensional torus is visualized for a pair of two pullers with $p=2.5$ in Fig.\,\ref{F3} by projecting the four-dimensional phase space of Eqs.\,(\ref{eq6}) onto a three-dimensional plane $(\Delta x,\Delta y,\Delta \phi=\phi_1-\phi_2)$. The initial conditions are chosen close to the orbitally unstable state. The trajectory appears to wrap around the unstable circular limit cycle, eventually covering the entire surface of a two-dimensional torus.
 
\begin{figure}[ht]
\includegraphics[width=0.9\hsize]{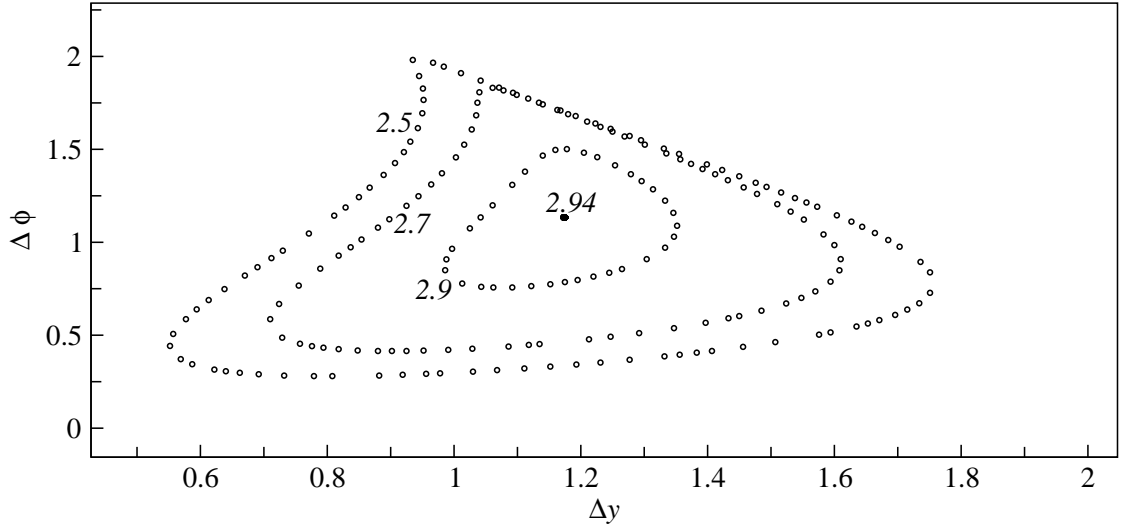} 
\caption{\label{F4} Poincar{\'e} sections $(\Delta x=0,\Delta y,\Delta \phi)$ for three different tori, found for two non-identical pullers with $p=2.5,2.7,2.9$ and $v=1$. An isolated point for $p=2.94$ corresponds to a stable circular limit cycle.} 
\end{figure}

In order to better understand the dynamics on a torus for two pullers, we construct a Poincar{\'e} section of the trajectories, using $(\Delta x=0, \Delta y>0,\Delta \phi)$ plane as a cross section, projected onto $(\Delta y,\Delta \phi)$. As shown in Fig.\,\ref{F4} for three different values of $p=2.5,2.7,2.9$, the smooth closed curves correspond to a motion on a two-dimensional torus.

It is well-known that continuation of invariant tori in a system of coupled differential equations is notoriously difficult \cite{Schilder05}. Standard numerical continuation method AUTO can be used to detect the birth of a torus, i.e. a torus bifurcation point, but it fails to continue the branch of the quasi-periodic solutions.  In order to overcome this problem we use a simple continuation protocol. Firstly, a torus is found by a direct integration of Eqs.\,(\ref{eq6}) with the initial conditions corresponding to the unstable periodic solution. Then the parameter $p$ is varied by a small increment $\pm 0.01$ and the new quasi-periodic solution is found with the initial conditions from the previous run. 

As a measure of the quasi-periodic states we choose the time average distance between swimmers $\bar{\rho}$
\begin{eqnarray}
\label{rhoav}
\bar{\rho}=\lim_{\bar{t}\rightarrow \infty}\frac{1}{\bar{t}}\int_{t_0}^{t_0+\bar{t}} \rho(t)\,dt,
\end{eqnarray}
where the moment of time $t_0$ is chosen sufficiently large so that the solution has reached a stable torus.
 We find two branches of stable invariant tori, born from the periodic solutions as a result of super-critical torus bifurcation at points $tr$ for $p<0.5$ and $p>2$. Characterised by the average distance $\bar{\rho}$  the branches are shown in Fig.\,\ref{F2} by thick solid lines. It is possible that each branch terminates at a saddle-node point, when two branches (one stable and one unstable) of invariant tori collide.  Clearly, only stable branch is accessible using the direct integration method.

\subsection{Stability diagram for a pair of two pushers and a pair of two pullers}
\label{stab_diag}
\begin{figure}[t]
\includegraphics[width=0.9\hsize]{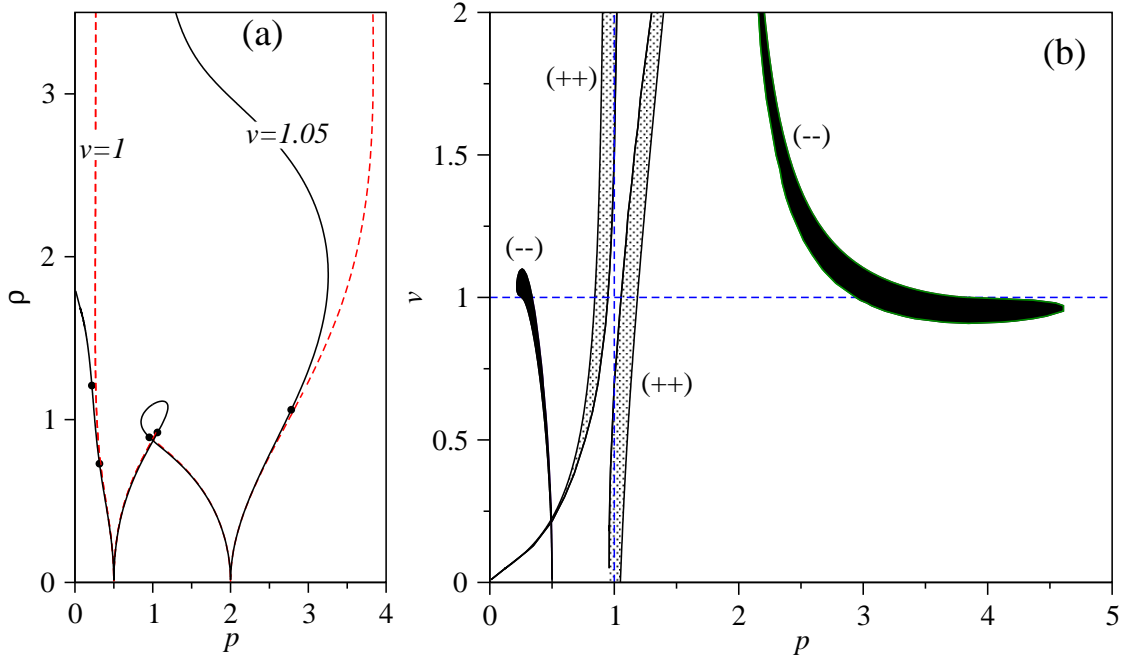} 
\caption{\label{F5} (a) Comparison of the bifurcation diagrams of periodic solutions for $v=1.05$ and $v=1$ with $p$ as a continuation parameter. For simplicity only torus bifurcation points are shown by filled circles for $v=1.05$. (b) Phase diagram of stable periodic solutions for a pair of two non-identical pushers (+\,+) and two pullers (-\,-) in the plane $(p,v)$. The diagram is invariant under simultaneous transformations $v\rightarrow 1/v$ and $p\rightarrow 1/p$.  } 
\end{figure}

Next, we study the effect of the mismatch between the self-propulsion speeds of the swimmers, quantified by the ratio $v=v_2/v_1$ on the existence of rotational equilibria. Interestingly, a slight variation of $v$ changes the bifurcation diagram of Eqs.\,(\ref{eq6}) dramatically. As an example, in Fig.\,\ref{F5}(a) we compute branches of periodic solutions of Eqs.\,(\ref{eq6}) for $v=1.05$ with $p$ as a principal continuation parameter. For comparison, the corresponding branches from Fig.\,\ref{F2}(a) are added. At $v=1.05$, the left branch ($p<0.5$) is no longer bounded by the saddle-node point, but extends all the way to $p=0$. This shows that when the self-propulsion speed of the swimmers is different, unstable periodic solutions can be found for arbitrary small (large) values of $p$. Indeed, in the scaling used here, if a certain rotational equilibrium (stable or unstable) is found for some values of $v$ and $p$, then the same equilibrium with identical stability must also exist for  $v\rightarrow 1/v$ and $p\rightarrow 1/p$. 
Using this symmetry, we conclude that if a branch of periodic solution exists for $v=1.05$ and some infinitesimaly small $p$, then  the same branch also exists for a very large $p\rightarrow 1/p$ and $v=1.05^{-1}=0.9524$.

For a better readability of the diagram in Fig.\,\ref{F5}(a) we only show the torus bifurcation points for $v=1.05$ (filled circles). Unlike for $v=1$, the left branch ($p<0.5$) for $v=1.05$ has two torus bifurcation points. The part of the left branch between the two torus points corresponds to sable periodic solutions for two pullers. As $v$ is varied and approaches the value of $v=1$ the left torus point becomes a saddle-node bifurcation point (sn) and the left branch extends to infinitely large values of $\rho$, as in Fig.\,\ref{F2}(a). The right branch ($p>2$) for $v=1.05$ has a similar structure as the right branch for $v=1$. 

In summary, we see that stable periodic solutions for two pullers exists either between a saddle-node and a torus bifurcation points, or between two torus bifurcation points. Next, we trace the saddle-node bifurcation points on the left and right branches and the torus bifurcation points on all three branches in the parameter plane $(p,v)$ and plot their loci in Fig.\ref{F5}(b). The resulting phase diagram shows the regions on the plane $(p,v)$, where stable periodic solutions exists for a pair of two non-identical pushers (++) and a pair of two pullers (-\,-). The diagram is invariant under simultaneous transformations $v\rightarrow 1/v$ and $p\rightarrow 1/p$.

\section{Three dimensional stability}
\label{3d}
As it was explained in Section\,\ref{orbits}, if the vectors ${\bm e}_1$, ${\bm e}_2$ and ${\bm r}_{12}$ are initially co-planar, i.e. if ${\bm r}_{12}\cdot ({\bm e}_1 \times {\bm e}_2)=0$, then the motion of the swimmers remains planar at all times. Therefore, the two-dimensional periodic and quasi-periodic orbits, found in Section\,\ref{orbits} also exist for two swimmers moving in the bulk of an unbounded fluid. However,  the stability of the orbits is significantly different with respect to three-dimensional perturbations of the swimmer's positions and orientations.

To study the stability of the planar orbits in three dimensions, we relax the condition $z=0$ and introduce the three dimensional relative position vector ${\bm \Delta r}=(x_1-x_2,y_1-y_2,z_1-z_2)$. Following \cite{michelin2010long}, we describe the orientation of each swimmer using the polar and the azimuthal angles $\theta_i$ and $\phi_i$, so that ${\bm e}_i=(\sin{\theta_i}\cos(\phi_i), \sin{\theta_i}\sin(\phi_i),\cos{\theta_i})$.

From Eqs.\,(\ref{eq2},\ref{eq3}) using the vector identity ${\bm a}\times ({\bm b}\times {\bm c})={\bm b}({\bm a}\cdot {\bm c})-{\bm c}({\bm a}\cdot {\bm b})$ we obtain for $i\not= k$
\begin{eqnarray}
\label{3d_eq1}
{\bm \Omega}_{ik}\times {\bm e}_k&=&\frac{3p_i}{4\pi\eta}\frac{({\bm e}_i\cdot {\bm r}_{ki})({\bm e}_i\times  {\bm r}_{ki})\times {\bm e}_k}{|{\bm r}_{ki}|^5}=\frac{3p_i}{4\pi\eta}\frac{({\bm e}_i\cdot {\bm \Delta r})( {\bm \Delta r} ({\bm e}_i \cdot {\bm e}_k)-{\bm e}_i ({\bm \Delta r}\cdot {\bm e}_k))}{|{\bm \Delta r}|^5}.
\end{eqnarray}

The time derivatives of the orientation vectors $d{\bm e}_i/dt$ are given by
\begin{eqnarray}
\label{3d_eq2}
\frac{d{\bm e}_i}{dt}=\left(\cos{\theta_i}\cos(\phi_i)\frac{d\theta_i}{dt}-\sin{\theta_i}\sin(\phi_i)\frac{d\phi_i}{dt},\cos{\theta_i}\sin(\phi_i)\frac{d\theta_i}{dt}+\sin{\theta_i}\cos(\phi_i)\frac{d\phi_i}{dt} ,-\sin{\theta_i}\frac{d\theta_i}{dt}\right).
\end{eqnarray}
The time derivatives of the angles $d\theta_i/dt$ and $d\phi_i/dt$ can be obtained from Eq.\,(\ref{eq3}), Eq.\,(\ref{3d_eq1}) and Eq.\,(\ref{3d_eq2}) as follows. From the $z$-component of Eq.\,(\ref{eq3}) we find
\begin{eqnarray}
\label{3d_eq3}
\frac{d\theta_i}{dt}&=&-\frac{1}{\sin{\theta_i}}\frac{3p_k}{8\pi\eta}\frac{({\bm e}_k\cdot {\bm \Delta r})( (z_1-z_2) ({\bm e}_1 \cdot {\bm e}_2)-\cos(\theta_k) ({\bm \Delta r}\cdot {\bm e}_i))}{|{\bm \Delta r}|^5}.
\end{eqnarray}
Next, eliminating $d\theta_i/dt$ from the $x$ and $y$ components of Eq.\,(\ref{3d_eq2})  we obtain 
\begin{eqnarray}
\label{3d_eq4}
\frac{d\phi_i}{dt}&=&\frac{1}{2\sin{\theta_i}}\left(\cos(\phi_i)({\bm \Omega_{ki}} \times {\bm e}_i)_y- \sin(\phi_i)({\bm \Omega_{ki}} \times {\bm e}_i)_x\right),
\end{eqnarray}
where the subscripts $x$ and $y$ denote the $x$ and $y$ components of the vector, respectively.

Finally, the relative three dimensional velocity is given by
\begin{eqnarray}
\label{3d_eq5}
\frac{d{\bm \Delta r}}{dt}=v_1{\bm e}_1-v_2{\bm e}_2+\frac{p_2}{8\pi\eta}\left(3\left(\frac{{\bm e}_2\cdot {\bm \Delta r}}{|{\bm \Delta r}|}\right)^2-1\right)\frac{{\bm \Delta r}}{|{\bm \Delta r}|^3}+\frac{p_1}{8\pi\eta}\left(3\left(\frac{{\bm e}_1\cdot {\bm \Delta r}}{|{\bm \Delta r}|}\right)^2-1\right)\frac{{\bm \Delta r}}{|{\bm \Delta r}|^3}.
\end{eqnarray}

Thus, the three-dimensional motion of two swimmers is described by a reduced seven-dimensional dynamical system. Comparing with Eqs.\,(\ref{eq6}), the system Eqs.\,(\ref{3d_eq3},\ref{3d_eq4},\ref{3d_eq5}) has three additional degrees of freedom: the relative $z$-coordinate $(z_1-z_2)$ and the two polar angles $\theta_i$.

The reduced system Eqs.\,(\ref{3d_eq3},\ref{3d_eq4},\ref{3d_eq5}) can be directly compared with the effective system of the far-field time-averaged equations, derived in \citet{michelin2010long} using multiple-scale analysis for the case of two swimmers with intrinsic rotation. Thus, after averaging over the fast time scale, associated with the intrinsic rotation of the orientation vectors, a four-dimensional dynamical system was derived in \citet{michelin2010long}, which describes the three-dimensional coupled motion of two {\it identical} force dipoles {\it without} self-propulsion. This situation corresponds to a special case of the reduced system derived here Eqs.\,(\ref{3d_eq3},\ref{3d_eq4},\ref{3d_eq5}), obtained by setting $v_i=0$ and $p_1=p_2$.

Eqs.\,(\ref{3d_eq3},\ref{3d_eq4},\ref{3d_eq5}) are nondimensionalized using scaling given in Section\,\ref{orbits}.  The two-dimensional periodic orbits in the seven-dimensional phase space are given by Eqs.\,(\ref{eq7}) extended by $\Delta z=0$ and $\theta_i=\pi/2$. Continuation method \cite{AUTO} is used to follow the branch off periodic orbits  using $p$ as the principal continuation parameter. All measures of the orbits $(\rho,\omega,R_i)$ are identical to those shown in Fig.\ref{F2}. The new information about the three-dimensional stability of the orbits is obtained by computing seven Floquet multipliers $\lambda_j$, $(j=1,\dots,7)$ of the Poincar{\'e} map linearized about the periodic orbit. 

Next we notice that the angles between three vectors ${\bm e}_i$ and ${\bm \Delta r}$ remain fixed at all times for a solution on a periodic orbit given by Eqs.\,(\ref{eq7}). This implies that such an orbit has all types of symmetries of a rigid body in three dimensional space. In other words, any periodic orbit is invariant under the perturbations along the orbit (phase shift) and the rotation of the plane of the orbit in three dimensional space of vector ${\bm \Delta r}$ in such a way that the angles between vectors ${\bm e}_i$ and ${\bm \Delta r}$ are preserved. Because the rotation of a rigid body in three dimensional space can be described by three Euler angles, we conclude that three Floquet multipliers of the linearized Poincar{\'e} map are always equal to +1, i.e. $\lambda_{1,2,3}=1$.  The remaining  four non-trivial multipliers determine the asymptotic stability of the solution. 

\begin{figure}[t]
\includegraphics[width=0.7\hsize]{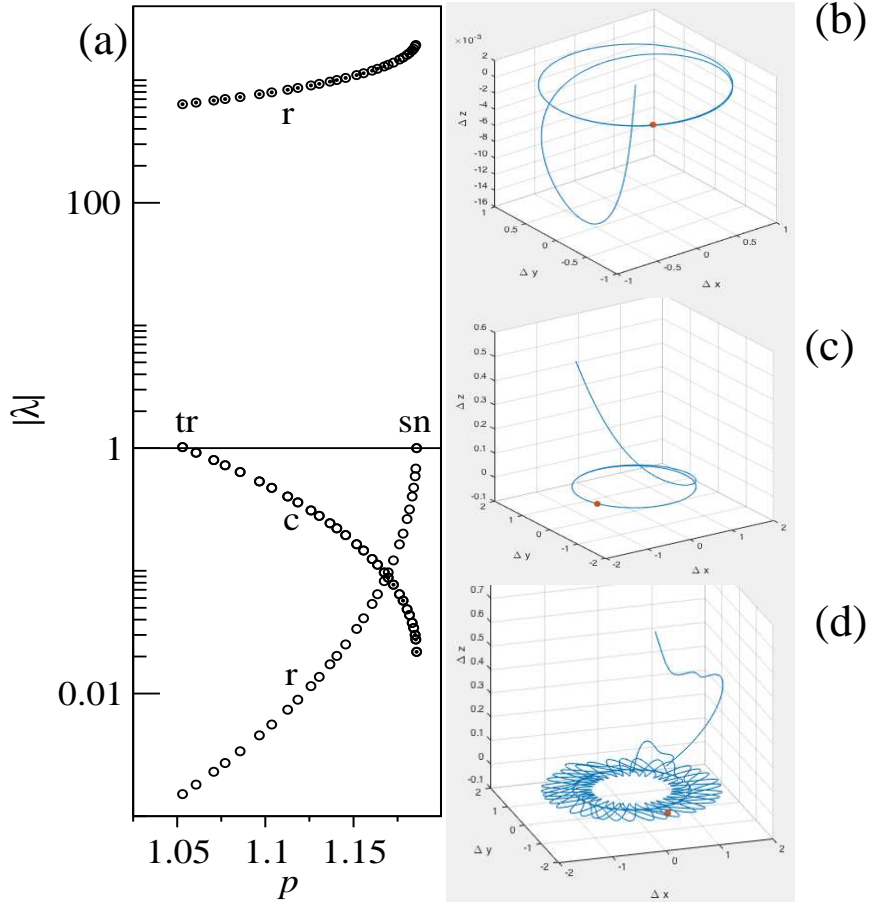} 
\caption{\label{F6} (a) Absolute values $|\lambda|$ of the four non-trivial Floquet multipliers of periodic orbits for a pair of two pushers on the branch between the torus point $p=1.05$ and the saddle-node bifurcation point $p=1.18$ in Fig.\ref{F2}. Complex (real) multipliers are labeled as "c" and "r". The orbit is stable with respect to two-dimensional perturbations and unstable with respect to three-dimensional perturbations. (b) Escape from the unstable orbit for a pair of two pushers with $p=1.1$ (c) Escape from the unstable orbit for a pair of two pullers at $p=3$. (d) Escape from the motion on a torus for a pair of two pullers with $p=2.7$. In (b,c,d) the trajectory corresponds to the solution of  Eqs.\,(\ref{3d_eq3},\ref{3d_eq4},\ref{3d_eq5}) with the initial conditions close to the corresponding periodic orbit. The initial point is marked by a circle.} 
\end{figure}

We find that periodic orbits for a pair of two pushers, or a pair of two pullers that are stable with respect to two-dimensional perturbations (solid lines in Fig.\,\ref{F2}(a,b)), are monotonically unstable with respect to three dimensional perturbations, as characterised by a single real unstable Floquet multiplier $|\lambda|>1$. In Fig.\ref{F6}(a) we choose to show the absolute values of the four non-trivial Floquet multipliers calculated for periodic orbits for a pair of two pushers with $v=1$ and parameter $p$ between the torus point $p=1.05$ and the saddle-node point $p=1.18$. The corresponding branch of solutions is shown by the solid line in Fig.\,\ref{F2}(b). Real (complex) multipliers are labeled as "r" ("c"). One real multiplier is always unstable $|\lambda|>1$, which gives rise to a monotonically developing three-dimensional perturbation. 

To visualize the dynamics of the unstable perturbation, we solve Eqs.\,(\ref{3d_eq3},\ref{3d_eq4},\ref{3d_eq5}) numerically with the initial conditions in the vicinity of the periodic orbit at $p=1.1$. Fig.\ref{F6}(b) shows the relative position vector ${\bm \Delta r}$ during the escape from the unstable orbit. After a very short time, the solution approaches the origin, i.e. ${\bm \Delta r}=0$, which corresponds to a collision of the pushers.
The escape from the unstable orbit for a pair of two pullers is also monotonic, however, the pullers run away from each other in the long-time limit, as shown in Fig.\ref{F6}(c) for the orbit found at $p=3$. Finally, in Fig.\ref{F6}(d) we demonstrate the escape dynamics from the motion on a torus for a pair of two pullers with $p=2.7$. The initial conditions are chosen close to the torus solution, as indicated by a circle. In all three cases, the escape from the planar equilibrium occurs monotonically via the direction, orthogonal to the plane of the equilibrium.

\section{Conclusions}
\label{conc}

We have studied the co-planar motion of two hydrodynamically coupled non-identical force-free swimmers, when the separation distance between the swimmers is much larger than their size. In this limit, the leading order flow induced by each swimmer is described by a force dipole term. If the orientations of the swimmers and their relative position vector are initially in one plane, then the motion remains co-planar at all times. 
Our results equally apply to the co-planar motion in the bulk of a fluid and at a stress-free liquid-gas interface.
For a two-dimensional motion, translational symmetry is used to reduce  the equations of motion to a four-dimensional dynamical system for the two-dimensional relative position vector and two orientation angles, as described by Eqs.\,(\ref{eq6}).  

We have found a closed analytical form of a rotational equilibrium, which corresponds to a steady circular orbital motion about a common origin with either identical or different radii of circular trajectories, as given by Eqs.\,(\ref{eq7}) . Any such circular orbital state is associated with periodic solutions of the reduced dynamical system modulo a phase gain of $2\pi$ for the orientation angles. Using numerical continuation method we have traced the branches of circular orbital states by varying the ratio of the dipole strengths $p=p_2/p_1$, or the ratio of self-propulsion velocities $v=v_2/v_1$.

 It was found that for a certain values of $p\not=1$ and (or) $v\not=1$, for a pair of two non-identical pushers $(p_i>0)$ or for a pair of two non-identical pullers $(p_i<0)$ the rotational equilibrium state is orbitally stable with respect to two-dimensional perturbations that preserve co-planer nature of vectors ${\bm e}_i$ and the relative position vector ${\bm r}_{ik}$. In the plane of parameters $(p,v)$ we present the universal stability diagram of the orbital states in Fig.\,(\ref{F5})(b). The universality of the diagram is reflected in its independence on the viscosity of the fluid as well as the characteristics of individuals swimmers $p_i$ and $v_i$. 

Following a branch of periodic orbits using either $p$ or $v$ as a continuation parameter, the stability of the orbit is lost or regained as the result of a torus, or a saddle-node bifurcation Fig.\,(\ref{F2}. In two dimensions, the unstable states may be either monotonically unstable (one real unstable Floquet multiplier), or oscillatory unstable (two complex unstable multipliers). 

For a pair of two non-identical pullers we find a new quasi-periodic planar equilibrium, which is born  from the periodic orbit as a result of a super-critical torus bifurcation. The quasi-periodic equilibrium is associated with the motion on a two-dimensional torus in the four-dimensional phase space. In the real space, the centre of mass of the pullers follows a circular trajectory, while the distance between the pullers oscillates with a new incommensurate frequency.

Previously, quasi-periodic rotational equilibria for the pair of hydrodynamically coupled micro swimmers have only been reported for the pair of two identical Quadroars, i.e. swimmers, whose bodies consist of four linked rotary discs \cite{Mirzakhanloo2018}. Numerical simulations revealed that these orbiting states are either long-lived transients, or truly stable equilibria. Earlier studies of two more simple model squirmers \cite{Ishikawa2006} or two spherical swimming cells with an intrinsic rotation \cite{michelin2010long} reported the existence of several different types of steady or periodic equilibria, however all of which were unstable in three dimensions.

Our results are in agreement with the previous study \cite{michelin2010long} for two identical swimmers in the far field. A pair of non-identical pushers or pullers may form a planar rotational equilibrium (periodic or quasi-periodic), which is stable with respect to two-dimensional perturbations. However, any such equilibrium was shown to be monotonically unstable with respect to three-dimensional perturbations. The swimmers escape from the unstable orbit by either colliding, as seen for two pushers, or separating and moving away from each other, as observed for two pullers.

Stability of the orbits in two dimensions presents an intriguing prospective of observing such equilibria in dilute suspensions of self-propelled micro-swimmers with distributed intrinsic characteristics, which are bound to move along a stress-free interface. Thus, our results show that a relatively small mismatch of the dipole strengths parameters $p_i$ of as little as 10\%, could be sufficient to observe a capture of two non-identical pushers into a stable periodic orbit at a planar liquid-gas interface.

\end{document}